# Super-Gaussian, super-diffusive transport of multi-mode active matter

*Seungsoo Hahn*[1,2], *Sanggeun Song*[1-3], *Dae Hyun Kim*[1-3], *Gil-Suk Yang*[1,3], *Kang Taek Lee*[4]*
*Jaeyoung Sung*[1-3]*

[1]Creative Research Initiative Center for Chemical Dynamics in Living Cells, Chung-Ang University, Seoul 06974, Korea.

[2]Department of Chemistry, Chung-Ang University, Seoul 06974, Korea.

[3]National Institute of Innovative Functional Imaging, Chung-Ang University, Seoul 06974, Korea.

[4]Department of Chemistry, Gwangju Institute of Science and Technology, Gwangju 61005, Korea

Corresponding Authors: J. Sung (jaeyoung@cau.ac.kr); K. T. Lee (ktlee@gist.ac.kr)

**Abstract**

Living cells exhibit multi-mode transport that switches between an active, self-propelled motion and a seemingly passive, random motion. Cellular decision-making over transport mode switching is a stochastic process that depends on the dynamics of the intracellular chemical network regulating the cell migration process. Here, we propose a theory and an exactly solvable model of multi-mode active matter. Our exact model study shows that the reversible transition between a passive mode and an active mode is the origin of the anomalous, super-Gaussian transport dynamics, which has been observed in various experiments for multi-mode active matter. We also present the generalization of our model to encompass complex multi-mode matter with arbitrary internal state chemical dynamics and internal state dependent transport dynamics.

Living cells in migration regulate their consumption of intracellular chemical energy according to the instructions encoded in their genes; they exhibit multiple transport modes during transport, consisting of two characteristic motions: a self-propelled, ballistic motion when the matter is in an active mode and an undirected, random motion when the matter is in a passive mode. Depending on the regulatory state of the cellular reaction networks underlying cell migration, the transport mode of living cells switches repeatedly between the active and the passive mode. This feature in living cell trajectories appears similar to that of Lévy walks [1,2]. Another interesting feature of living cells' motion is that they repeatedly reverse their direction. This run-and-reverse motion has been reported in various bacterial systems [3-7]. These features have also been observed in the transport of various types of cargos and vesicles in living cells [8-10]. Active matter, such as living cells and intracellular active particles, generally exhibits an anomalous, non-Gaussian transport dynamics, which cannot be described by Einstein's theory of Brownian motion [11,12] or more recent theories for anomalous transport in a disordered environment [13-17].

There are models of passively moving particles that have been used to explain the long time behavior of the mean square displacement (MSD) of multi-mode active matter observed in experiments [18,19]. Although these models assume that the stochastic dynamics of multi-mode active matter is qualitatively the same as that of passive matter, they are able to provide a satisfactory explanation of experimental results for the long time behavior of the MSD in many cases [20,21]. However, experimental data with a higher time resolution revealed that multi-mode active matter has qualitatively different stochastic dynamics from passive matter; the MSD of multi-mode active matter shows short time-diffusive motion, intermediate super-diffusive motion, and long-time diffusive motion with a greater diffusion coefficient [22,23], which cannot be explained by the passive matter models [24,25]. An alternative model to account for the anomalous transport dynamics of active matter is the "active Brownian particle" model. In this model, a velocity-dependent friction in the Langevin equation is used to describe the self-propelled motion of active particles [7,26]. The active Brownian particle model does provide an enhanced explanation for the anomalous MSD of active particles; however, this model and the models mentioned above cannot explain the anomalous displacement distribution of active matter, whose spatial distribution is non-Gaussian with a positive excess kurtosis [6,27]. A number of other interesting models have been proposed for self-propelled particles [28-34]. However, to the best of our knowledge, none of them represents multi-mode active matter, which switches between an active, self-propelled transport mode and a seemingly passive, random mode depending on its internal state dynamics.

In this Letter, we present an exactly solvable model for the stochastic transport of multi-mode active matter. In the high friction regime, where we can safely neglect the inertial term in the Langevin equation, the velocity $\dot{x}(t)$ of multi-mode active matter with a friction constant, $\gamma$, can be written as the sum of two components:

$$\dot{x}(t) = v_s(\Gamma(t)) + \gamma^{-1}\xi(t), \tag{1}$$

where $v_s(\Gamma)$ and $\gamma^{-1}\xi(t)$ represent the velocity component of a self-propelled, ballistic motion, which is dependent on the internal state $\Gamma$ and the velocity component caused by the random fluctuating force. Assuming that the dynamics of the random fluctuating force occurs in a time scale far shorter than the internal state dynamics, we model $\xi(t)$ as Gaussian white noise, whose time correlation, $\langle \xi(t+t_0)\xi(t_0) \rangle$, is proportional to the Dirac delta function, $\delta(t)$. On the other hand, the relaxation of $\langle v_s(\Gamma(t+t_0))v_s(\Gamma(t_0))\rangle$ from the initial value, $\langle v_s^2 \rangle$, to the final value, $\langle v_s \rangle^2$, occurs in the time scale of the internal state dynamics. We assume that the cell state dynamics is an arbitrary stochastic process that can be represented by a multidimensional Markov process. The Fokker-Planck equation corresponding to Eq. (1) is given by

$$\frac{\partial}{\partial t}P(\Gamma,x,t) = \frac{\partial}{\partial x}\left[D_0\frac{\partial}{\partial x} - v_s(\Gamma)\right]P(\Gamma,x,t) + L(\Gamma)P(\Gamma,x,t), \qquad (2)$$

where $P(\Gamma,x,t)$ denotes the probability density function (PDF) of active matter with the position $x$ and internal state $\Gamma$ at time $t$ [35,36]. In Eq. (2), $D_0$ stands for the diffusion coefficient for passive motion originating from the random fluctuating force, which is defined by $D_0 = \gamma^{-2}\int_0^\infty dt\langle \xi(t)\xi(0)\rangle$. $L(\Gamma)$ denotes the mathematical operator describing the internal state dynamics of the system. Our model yields analytic results for the MSD and the non-Gaussian parameter [37].

Here, we compare two simple, exactly solvable models of active matter: one for single-mode active matter and the other for multi-mode active matter. These models are shown in Fig. 1. For the single-mode model, which only exhibits an active mode, as shown in Fig. 1(a), Eq. (2) yields

$$\frac{\partial}{\partial t}\begin{pmatrix} P_+(x,t) \\ P_-(x,t) \end{pmatrix} = \frac{\partial}{\partial x}\left[\frac{\partial}{\partial x}D_0\mathbf{I} - \begin{pmatrix} v_a & 0 \\ 0 & -v_a \end{pmatrix}\right]\begin{pmatrix} P_+(x,t) \\ P_-(x,t) \end{pmatrix} + \begin{pmatrix} -k_a & k_a \\ k_a & -k_a \end{pmatrix}\begin{pmatrix} P_+(x,t) \\ P_-(x,t) \end{pmatrix}. \qquad (3)$$

For the multi-mode model, which exhibits both active and passive modes, as shown in Fig. 1(b), Eq. (2) yields

$$\frac{\partial}{\partial t}\begin{pmatrix} P_+(x,t) \\ P_0(x,t) \\ P_-(x,t) \end{pmatrix} = \frac{\partial}{\partial x}\left[\frac{\partial}{\partial x}D_0\mathbf{I} - \begin{pmatrix} v_a & 0 & 0 \\ 0 & 0 & 0 \\ 0 & 0 & -v_a \end{pmatrix}\right]\begin{pmatrix} P_+(x,t) \\ P_0(x,t) \\ P_-(x,t) \end{pmatrix} + \begin{pmatrix} -k_0 & k_a & 0 \\ k_0 & -2k_a & k_0 \\ 0 & k_a & -k_0 \end{pmatrix}\begin{pmatrix} P_+(x,t) \\ P_0(x,t) \\ P_-(x,t) \end{pmatrix}. \quad (4)$$

In Eqs. (3) and (4), $P_i(x,t)$ designates the probability density of the matter at state $\Gamma_i$ ($i \in +, -, 0$) and position $x$ at time $t$. $\pm v_a$, $k_a$, and $k_0$ denote, respectively, the velocity $v_s(\Gamma_\pm)$ of the self-propelled motion of the matter at state $\Gamma_\pm$, the transition rate to either state, $\Gamma_+$ or $\Gamma_-$, of the matter in active mode, and the transition rate to the state, $\Gamma_0$, of the matter in passive mode, at which $v_s(\Gamma_0) = 0$. Typical time traces are displayed for the two different models in Fig. 1 and Supplemental Material [38]. Exact analytic solutions of Eqs. (3) and (4) can be obtained in the Fourier domain [37].

From the exact solutions, we obtain the distribution $f(\bar{v},t)$ of the mean velocity, defined by $x(t)/t (\equiv \bar{v}(t))$; it is shown in Fig. 2 for each model. In the short-time limit, for both models, the mean velocity distribution is found to be a linear combination of Gaussians centered at the state-dependent self-propelled velocity, $v_s(\Gamma_i)$, that is,

$$f(\bar{v},t) \cong \sum_i p_i^{eq} G(\bar{v} - v_s(\Gamma_i), 2D_0/t) \qquad (t \ll \tau_c), \quad (5)$$

with $p_i^{eq}$ being the equilibrium probability of state $\Gamma_i$, given by $p_\pm^{eq} = 1/2$ for the single-mode model and by $p_\pm^{eq} = k_a/(k_0 + 2k_a)$ and $p_0^{eq} = k_0/(k_0 + 2k_a)$ for the multi-mode model. In Eq. (5), $G(v, \sigma^2)$ denotes the Gaussian distribution of $v$ with the mean and variance given by 0 and $\sigma^2$, respectively. Equation (5) can also be obtained from the distribution of the instantaneous velocity given in Eq. (1), because the mean velocity, $x(t)/t$, is the same as the instantaneous velocity in the short-time limit [37]. Equation (5) serves as a good approximation of the mean velocity distribution at moderately short times, where a transition between states has yet to occur, or at times earlier than the characteristic relaxation time, $\int_0^\infty dt \phi_{v_s}(t) [\equiv \tau_c]$, where $\phi_{v_s}(t)$ denotes the normalized time correlation function, $\langle v_s(t) v_s(0) \rangle / \langle v_s^2 \rangle [\equiv \phi_{v_s}(t)]$, of the internal state dependent self-propelled velocity, $v_s(\Gamma)$.

The long-time distribution of the mean velocity becomes Gaussian for both single-mode and multi-mode models of active matter, while the short-time distribution can vary depending on the model. At short times ($t \ll \tau_c$), the mean velocity distribution, $f_S(\bar{v},t)$, of the single-mode model has two Gaussian peaks centered at $+v_a$ and $-v_a$, which are the two velocities of self-propelled motion. In comparison, the mean velocity distribution, $f_M(\bar{v},t)$, of the multi-mode model at short times has an additional Gaussian peak centered at 0, resulting from the state, $\Gamma_0$, of the active matter in passive mode. The variance of each Gaussian peak, which originates from the random fluctuating force, is approximately given by $2D_0/t$. However, as shown in Fig. 2, both $f_M(\bar{v},t)$ and $f_S(\bar{v},t)$ converge to a Gaussian with a mean of zero and a variance proportional to $t^{-1/2}$ at long times [37]. That is to say, for both models, the distribution of $x(t)/\sqrt{t}$ approaches a Gaussian stable distribution at long times, in accordance with the Gaussian central limit theorem [37].

The relaxation dynamics of the mean velocity distribution is highly dependent on the characteristic relaxation time, $\tau_c$, of internal state dependent self-propelled velocity, $v_s(\Gamma)$. The mean velocity distribution approaches the long-time asymptotic Gaussian faster as the value of $\tau_c$ decreases (see Figs. 2(c) and 2(d)). The analytic expression of $\tau_c$ is dependent on the model in question. For the single-mode model, $\tau_c$ is given by half the lifetime, $k_a^{-1}$, of the state, $\Gamma_\pm$, of the active matter in active mode, i.e., $\tau_c = (2k_a)^{-1}$. For the multi-mode model, $\tau_c$ is the same as the lifetime, $k_0^{-1}$, of the state $\Gamma_\pm$ [37]. In the small $\tau_c$ limit, the mean velocity distribution is Gaussian at any finite time. Note also that the variance in the mean velocity, or the mean squared velocity, at any given time decreases with the relaxation speed, $\tau_c^{-1}$, of the fluctuation in the self-propelled velocity, as shown in Fig. 2(e). This is a common feature of dynamically disordered systems; in spectroscopy, it has been termed motional narrowing.

The mean velocity distribution, $f_M(\overline{v},t)$, of the multi-mode model is dependent on the lifetime, $\tau_0 (\equiv 1/2k_a)$, of passive mode, $\Gamma_0$, as well as on the lifetime, $\tau_a (\equiv 1/k_0 = \tau_c)$, of the states in active mode, $\Gamma_\pm$. As shown in Fig. 2(f), when the population ratio, $R(\equiv p_0^{eq}/(p_+^{eq} + p_-^{eq}) = \tau_0/\tau_a = k_0/(2k_a))$, of the state in passive mode to the states in active mode decreases, $f_M(\overline{v},t)$ approaches $f_S(\overline{v},t)$ [37]. However, as the value of $R$ increases, the peak centered at $\overline{v} = 0$ in $f_M(\overline{v},t)$ grows large, so that the MSD of the multi-mode model is smaller than the MSD of the single-mode model.

For both the models, the MSD has three different kinetic phases: the short-time diffusion phase, an intermediate super-diffusive phase, and the long-time diffusive phase with a greater diffusion coefficient, in agreement with the previous experimental results [22,23]. Exact analytic expressions of the MSD for both models can be written in the same formula,

$$\langle x^2(t)\rangle = 2(D_0 + D_a)t + 2D_a\tau_c\left(e^{-t/\tau_c} - 1\right), \tag{6}$$

where $D_a$ is the effective diffusion coefficient component contributed from the self-propelled motion, defined by $D_a \equiv \int_0^\infty dt \langle v_s(t)v_s(0)\rangle = \tau_c v_a^2 p_a$. Here, $p_a$ designates the probability of the states in active mode, which is given by unity for the single-mode model and by $p_a = p_+^{eq} + p_-^{eq} = (1+R)^{-1}$ for the multi-mode model. $D_0$ and $\tau_c$ have the same meaning as above. As shown in Fig. 3(a), the MSD is given by $\langle x^2(t)\rangle \cong 2D_0 t$ at short times ($t \ll \tau_c$) and dominantly contributed from the seemingly passive, random motion. On the other hand, at long times ($t \gg \tau_c$), the MSD is given by $\langle x^2(t)\rangle \cong 2(D_0 + D_a)t$, with the diffusion coefficient increased by $D_a$. In intermediate times ($2D_0/v_a^2 < t \ll \tau_c$), the MSD shows a super-diffusive behavior ($MSD \sim t^\alpha$ with $1 < \alpha \leq 2$). In the early stage of the intermediate region, the MSD is approximately a quadratic function of time, i.e., $\langle x^2(t)\rangle \cong 2D_0 t + D_a \tau_c (t/\tau_c)^2$, shown by the green lines in Fig. 3(b), which originates from the ballistic, self-propelled motion of active matter.

While both the single-mode and multi-mode models yield qualitatively the same analytic result for the MSD, the results they yield for the displacement distribution can be quite different from each other. The displacement distribution, $P_M(x,t)$, of the multi-mode model can be super-Gaussian, in accordance with the experimental data reported in Refs. [34,39], whereas $P_S(x,t)$ of the single-mode model is always sub-Gaussian. For the multi-mode model, the deviation of $P_M(x,t)$ of the multi-mode model from Gaussian measured by the non-Gaussian parameter, $\alpha_R(t)\left[\equiv \langle x^4(t)\rangle / \left(3\langle x^2(t)\rangle^2\right) - 1\right]$, is sensitive to the population ratio, $R$, of the state in passive mode to the states in active mode, which is shown in Fig. 3(c). The exact analytic expression of $\alpha_R(t)$ is presented in the Supplemental Material [37]. The simpler asymptotic expression of $\alpha_R(t)$ at both short times and long times is given by

$$\alpha_R(t) \cong \begin{cases} \dfrac{1}{12}\dfrac{(R-2)}{(R+1)^2}\left(\dfrac{D_a^{(0)}}{D_0}\right)^2 (t/\tau_c)^2, & t \ll \tau_c, \\ 2\dfrac{R^2-R-1}{(R+1)^3}\left(\dfrac{D_a^{(0)}}{D_0+D_a}\right)^2 \dfrac{\tau_c}{t}, & t \gg \tau_c, \end{cases} \quad (7)$$

where $D_a^{(0)}$ designates $\tau_c v_a^2$, or the value of $D_a$ in the limit where the state of active matter is always in the active mode. According to Eq. (7), the displacement distribution, $P_M(x,t)$, of multi-mode active matter is super-Gaussian when $R > 2$, but sub-Gaussian when $R < \left(1+\sqrt{5}\right)/2 \cong 1.62$ at all times [37]. However, when $\left(1+\sqrt{5}\right)/2 < R < 2$, the displacement distribution, $P_M(x,t)$, of multi-mode active matter switches from sub-Gaussian to super-Gaussian over time [37]. Note that $\alpha_R(t)$ vanishes in the large $R$ limit, where the state of multi-mode matter is always in passive mode. This means that, in our model, it is the self-propelled, ballistic motion that causes the displacement distribution to be non-Gaussian. In the opposite, small $R$ limit, $P_M(x,t)$ has exactly the same shape as $P_S(x,t)$ [37]. Thus, the non-Gaussian parameter, $\alpha_R(t)$, of the multi-mode model reduces to $\alpha_0(t)\left[=\lim_{R\to 0}\alpha_R(t)\right]$ of the single-mode model, whose asymptotic behavior is given by

$$\alpha_0(t) \cong \begin{cases} -\dfrac{1}{6}\left(\dfrac{D_a^{(0)}}{D_0}\right)^2 (t/\tau_c)^2, & t \ll \tau_c. \\ -2\left(\dfrac{D_a^{(0)}}{D_0 + D_a^{(0)}}\right)^2 \dfrac{\tau_c}{t}, & t \gg \tau_c. \end{cases} \qquad (8)$$

Equations (8) shows that the displacement distribution, $P_S(x,t)$, of single-mode active matter is sub-Gaussian only [37].

Both $P_M(x,t)$ and $P_S(x,t)$ approach Gaussian at long times; however, their deviation from Gaussian, which is measured by the non-Gaussian parameter, slowly decreases with time, following $t^{-1}$ at long times ($t \gg \tau_c$), according to Eqs. (7) and (8). As shown in Fig. 3(c), the deviation of the displacement distribution from Gaussian can be sizable even at long times where the MSD, given in Eq. (6), is linearly proportional to time. This has been observed, for example, in liposome diffusion in a nematic solution of actin filaments [40].

The multi-mode active matter model discussed above can be extended to a more complex model in the higher spatial dimension, $d$. For the generalized model, the stochastic differential equation corresponding to Eq. (1) is given by

$$\dot{\mathbf{r}}(t) = \mathbf{v}_s(\Gamma(t)) + \gamma^{-1}\boldsymbol{\xi}(t), \qquad (9)$$

where each bold symbol denotes the $d$-dimensional vector corresponding to each scalar quantity in Eq. (1). The general expression of the MSD obtained from Eq. (9) is given by

$$\langle |\mathbf{r}(t)|^2 \rangle = 2d \int_0^t d\tau (t-\tau) \left[ D_0 \tau_p^{-1} \phi_\xi(\tau) + D_a \tau_c^{-1} \phi_{\mathbf{v}_s}(\tau) \right], \tag{10}$$

where $D_0$, $\tau_p$, $D_a$, and $\tau_c$ are, respectively, defined by $D_0 = d^{-1} \gamma^{-2} \int_0^\infty dt \langle \xi(t) \cdot \xi(0) \rangle$, $\tau_p \equiv \int_0^\infty dt\, \phi_\xi(t)$, $D_a = d^{-1} \int_0^\infty dt \langle \mathbf{v}_s(t) \cdot \mathbf{v}_s(0) \rangle$, and $\tau_c \equiv \int_0^\infty dt\, \phi_{\mathbf{v}_s}(t)$. Here, $\phi_\mathbf{x}(t)$ denotes the normalized time correlation function, $\langle \mathbf{x}(t) \cdot \mathbf{x}(0) \rangle / \langle \mathbf{x}(0)^2 \rangle$, of vector $\mathbf{x}(t)$. The functional form of $\phi_{\mathbf{v}_s}(\tau)$ varies depending on the internal state dynamics and its coupling to the self-propelled velocity. Given that the relaxation time of random fluctuation force $\xi(t)$ is far shorter than the observation time $t$, Eq. (10) reduces to $\langle |\mathbf{r}(t)|^2 \rangle \cong 2dD_0 t + 2dD_a \tau_c^{-1} \int_0^t d\tau (t-\tau) \phi_{\mathbf{v}_s}(\tau)$. This result is the generalization of equation (6) for multi-dimensional systems with arbitrary $\phi_{\mathbf{v}_s}(\tau)$; it reduces to equation (6) for the one-dimensional model with $\phi_{\mathbf{v}_s}(\tau) = \exp(-t/\tau_c)$. In addition, the general expression of the non-Gaussian parameter can also be obtained from equation (9) as follows:

$$\alpha_R(t) = \frac{\langle |\mathbf{r}_{\mathbf{v}_s}(t)|^2 \rangle^2}{\langle |\mathbf{r}(t)|^2 \rangle^2} \alpha_{\mathbf{v}_s}(t). \tag{11}$$

Here $\alpha_{\mathbf{v}_s}(t)$ is defined by $\alpha_{\mathbf{v}_s}(t) \equiv \frac{d}{d+2} \langle |\mathbf{r}_{\mathbf{v}_s}(t)|^4 \rangle / \langle |\mathbf{r}_{\mathbf{v}_s}(t)|^2 \rangle^2 - 1$ with $\langle |\mathbf{r}_{\mathbf{v}_s}(t)|^2 \rangle$ and $\langle |\mathbf{r}_{\mathbf{v}_s}(t)|^4 \rangle$ defined as $\int_0^t \int_0^t d\tau_2 d\tau_1 \langle \mathbf{v}_s(\tau_2) \cdot \mathbf{v}_s(\tau_1) \rangle$ and $\int_0^t \int_0^t \int_0^t \int_0^t d\tau_4 d\tau_3 d\tau_2 d\tau_1 \langle \mathbf{v}_s(\tau_4) \cdot \mathbf{v}_s(\tau_3) \mathbf{v}_s(\tau_2) \cdot \mathbf{v}_s(\tau_1) \rangle$, respectively. The non-Gaussian parameter given in Eq. (11) vanishes in both the short time and the long time limits. At times

far shorter than the relaxation time scale, $\tau_c$, of the self-propelled velocity, $\left(\left\langle\left|\mathbf{r}_{\mathbf{v}_s}(t)\right|^2\right\rangle\Big/\left\langle\left|\mathbf{r}(t)\right|^2\right\rangle\right)^2$, and hence the non-Gaussian parameter given in Eq. (11), vanish [37]. On the other hand, in the long time limit, $\left(\left\langle\left|\mathbf{r}_{\mathbf{v}_s}(t)\right|^2\right\rangle\Big/\left\langle\left|\mathbf{r}(t)\right|^2\right\rangle\right)^2$ approaches $D_a/(D_0+D_a)$ but $\alpha_{\mathbf{v}_s}(t)$, or the non-Gaussian parameter of the self-propelled displacement, $\int_0^t d\tau \mathbf{v}_s(\tau)$, vanishes because the distribution of the self-propelled displacement becomes Gaussian according to the Gaussian central limit theorem. However, the non-Gaussian parameter has a non-zero value between the two limits. In the simple one-dimensional multi-mode active matter model with the Poisson state switching dynamics, we can show that equation (11) reduces to equation (7) [37]. Equations (10) and (11) enable us to calculate the MSD and non-Gaussian parameter for general multi-mode active matter with possibly non-Poisson state switching dynamics.

In summary, we present an analytic theory and an exactly solvable model of multi-mode active matter, which switches between an active, self-propelled transport mode and a seemingly passive, random mode depending on its internal state chemical dynamics. Our exact model study clearly shows that the reversible transition between seemingly passive, random motion and the self-propelled, ballistic motion is an important source of the super-Gaussian displacement distribution commonly observed for multi-mode active matter. This model is sufficiently flexible so that it can be easily generalized to encompass multi-state, multi-mode active matter with arbitrary internal state chemical dynamics and internal state coupled transport dynamics. The application of the present approach to the quantitative explanation of experimental results for examples of multi-mode active matter is to be published elsewhere.

FIGURES

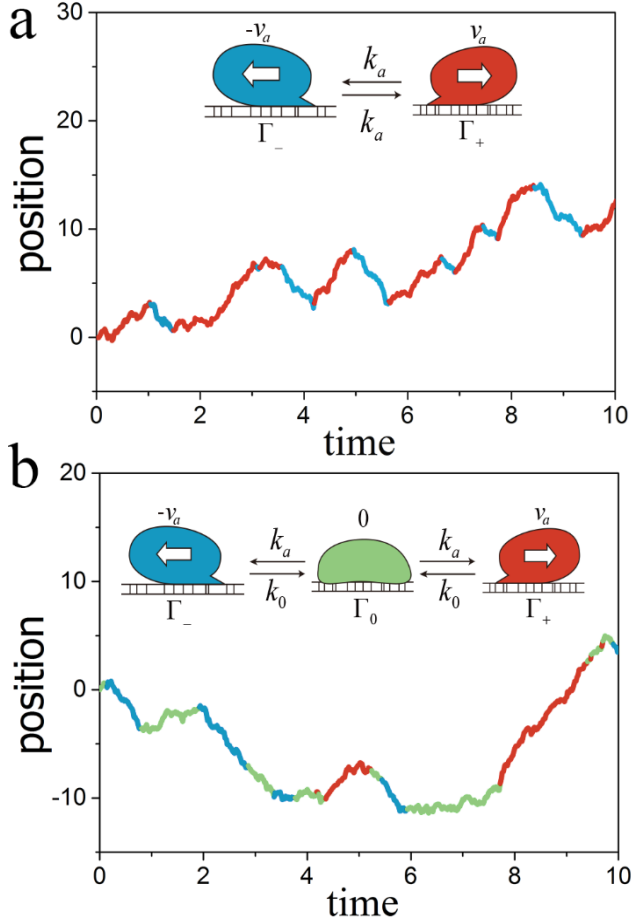

**FIG. 1. Model systems and typical trajectories.** (a) The single-mode model consists of two internal states, $\Gamma_+$ and $\Gamma_-$. The single-mode active matter in $\Gamma_\pm$ state performs self-propelled, directed motion with velocity $\pm v_a$ under a random fluctuating force exerted from medium. The stochastic transition between internal states is characterized by the rate constant, $k_a$. (b) The multi-mode model consisting of three internal states: passive transport state, $\Gamma_0$, in addition to active transport states, $\Gamma_+$ and $\Gamma_-$. The multi-mode matter performs undirected, random motion in state $\Gamma_0$, but performs directed, self-propelled motion with velocity $\pm v_a$ in state $\Gamma_\pm$. $k_a$ and $k_0$ represent the stochastic transition rates from the passive $\Gamma_0$ to the active $\Gamma_\pm$ state and from the active $\Gamma_\pm$ to the passive $\Gamma_0$ state, respectively. For each model, a typical time trace of the position is shown. Colors in the active matter diagram and trajectory represent the cell's internal states.

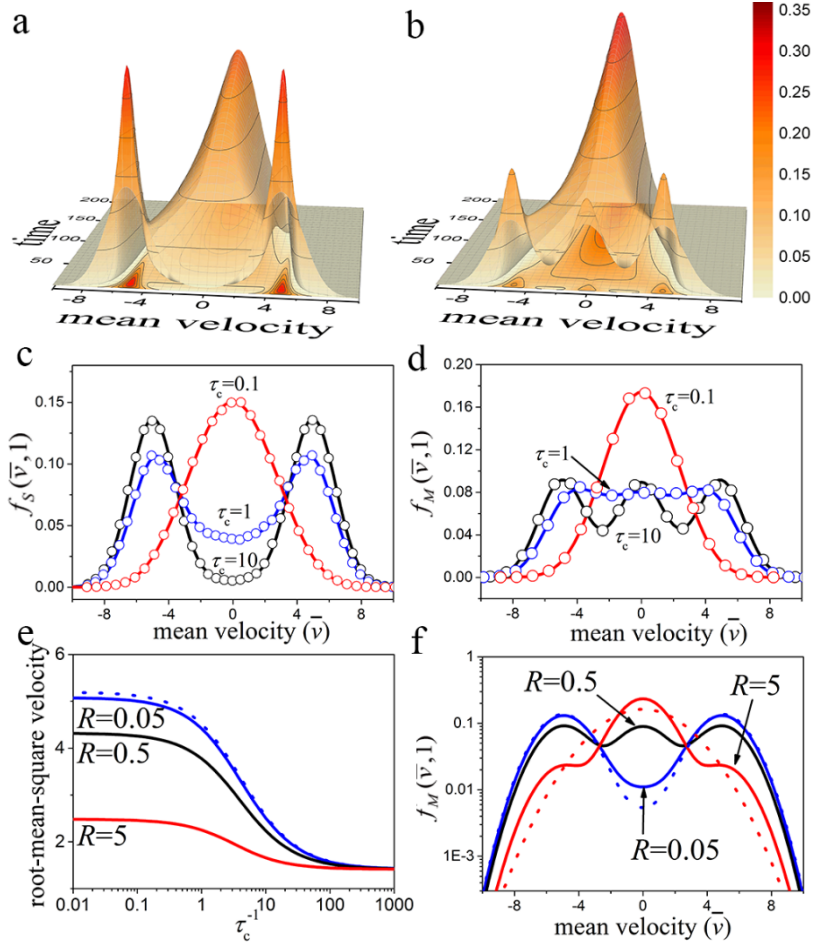

**FIG. 2. PDFs for mean velocity distribution.** The time dependent mean velocity distribution, (a) $f_S(\bar{v},t)$ for the single-mode model and (b) $f_M(\bar{v},t)$ for the multi-mode model with $R=0.5$. In both (a) and (b), the mean velocity distribution is displayed starting from arbitrary unit time, $T_u$, and the relaxation time $\tau_c$ of the velocity-velocity auto-correlation function is set to be $10T_u$. The mean velocity distribution at $t=T_u$ (c) for the single-mode model and (d) for the multi-mode model, with three different values of $\tau_c$. (lines) analytic results (circles) stochastic simulation results. In (d), the value of $R$ is set to be $0.5$, in which case the three states are equally probable at equilibrium. (e) Dependence of the root-mean-square velocities, or the standard deviation of the mean velocity distributions on the relaxation speed measured by $\tau_c^{-1}$, and (f) the mean velocity distribution at $t=T_u$ for the multi-mode model with three different values of $R$: (blue dotted line) $R=0$; (blue solid line) $R=0.05$; (black line) $R=0.5$; and (red line) $R=5$. The value of $\tau_c$ is $10T_u$. In the small $R$ limit, $f_M(\bar{v},1)$ approaches $f_S(\bar{v},1)$. The Gaussian distribution with the same mean and variance as $f_M(\bar{v},1)$ for $R=5$ is plotted as a red dotted line. The values of the other parameters are set to be $D_0=1$ and $v_a=5$ for all cases.

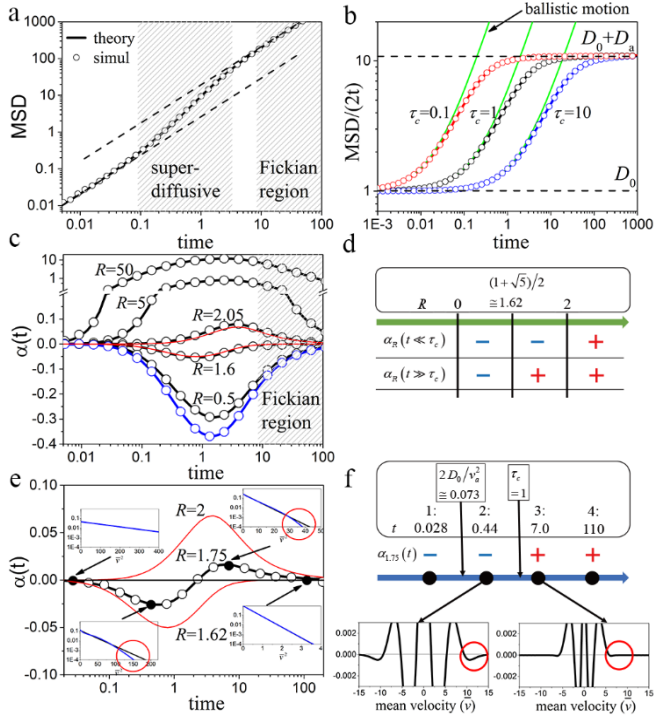

**FIG. 3. Mean square displacement and non-Gaussian parameter.** (a) Time-dependent mean square displacement (MSD). The single-mode and multi-mode models share the same MSD, given in Eq. (6). The value of $\tau_c$ is set to 1. The values of the other parameters, $D_0$ and $D_a$ are set to be $D_0 = 1$ and $D_a = 10$. (lines) analytic results (circles) stochastic simulation results. (b) Dependence of $\langle x^2(t) \rangle / 2t$ on time for the three cases with $\tau_c = 0.1$ (red), $\tau_c = 1$ (black), and $\tau_c = 10$ (blue). The effective diffusion coefficient increases from $D_0$ to $D_0 + D_a$, whose transition time scale is determined by $\tau_c$. The green line represents the ballistic motion ( $\langle x^2(t) \rangle \cong 2D_0 t + D_a \tau_c (t/\tau_c)^2$ ) corresponding to each case. (c) The non-Gaussian parameter, $\alpha(t)$, for the single-mode model (blue line) and for the multi-mode models with various values of $R$ (black lines). The two red lines represent $\alpha_M(t)$ for the two critical values of $R$, 1.62 and 2. (d) $\alpha_M(t)$ is always negative when $R < (1+\sqrt{5})/2 \cong 1.62$, but positive at all times when $R > 2$. When $1.62 < R < 2$, $\alpha_M(t)$ switches from a short-time negative regime to a long-time positive regime. (e) As a representative case for the time-dependent switching, $\alpha_M(t)$ with $R = 1.75$ is plotted as a black line. The two red lines represent $\alpha_M(t)$ for the two critical values of $R$, shown in (c). The non-Gaussian parameter vanishes both in the short time and the long time limits, meaning that the initial distribution is a delta function, Gaussian with zero variance, and the final distribution obeys the Gaussian central limit theorem. At the four time points marked by the solid circles, the mean velocity

distributions (blue line) and their corresponding Gaussian distributions (black line) are plotted according to the square mean velocity in the insets. The red circles mark the deficiency in population of the active matter in the high mean velocity region, compared to their corresponding Gaussian distribution. (f) Deviation of the mean velocity distribution from Gaussian at the two time points marked by the filled black circles. The red circles here and in the insets in (e) both represent the same spatial regime.

**Supplemental Material for**

**"Super-Gaussian, super-diffusive transport of multi-mode active matter"**

Seungsoo Hahn,[2] Sanggeun Song,[1,2,3] Dae Hyun Kim,[1,2,3] Gil-Suk Yang,[1,2,3] Kang Taek Lee,[4] Jaeyoung Sung[1,2,3,†]

[1]Creative Research Initiative Center for Chemical Dynamics in Living Cells, Chung-Ang University, Seoul 06974, Korea.

[2]Department of Chemistry, Chung-Ang University, Seoul 06974, Korea.

[3]National Institute of Innovative Functional Imaging, Chung-Ang University, Seoul 06974, Korea.

[4]Department of Chemistry, Gwangju Institute of Science and Technology, Gwangju 61005, Korea

**Contents**

## A. Derivation of the second and fourth moments of displacement for the multi-mode model

The multi-mode active matter model has three internal states, $\Gamma_+$, $\Gamma_0$, and $\Gamma_-$. Each internal state regulates the direction and speed of a given active matter as explained in Fig. 1(b). Based on the initial conditions that the internal states are initially in equilibrium and the initial position of active matter is zero, three simultaneous equations are obtained from Eq. (4) by applying the Fourier transform and the Laplace transform to $P_i(x,t)$ with $i \in +, 0, -$. The solution of the simultaneous equations provides three probability density functions (PDF) for the individual internal states in the Fourier-Laplace domain, written as

$$\begin{pmatrix} \tilde{P}_+(w,s) \\ \tilde{P}_0(w,s) \\ \tilde{P}_-(w,s) \end{pmatrix} = \frac{1}{(s+D_0w^2)(\chi(w,s)+v_a^2w^2)+2k_a v_a^2 w^2} \begin{pmatrix} p_+^{eq}(\chi(w,s) - i v_a w(s+D_0w^2+k_0+2k_a)) \\ p_0^{eq}(\chi(w,s) + v_a^2 w^2) \\ p_-^{eq}(\chi(w,s) + i v_a w(s+D_0w^2+k_0+2k_a)) \end{pmatrix}$$

with $\chi(w,s) \equiv (s+D_0w^2+k_0)(s+D_0w^2+k_0+2k_a)$. (S1)

In Eq. (S1), $w$ and $s$ respectively denote the Fourier transform of position $x$ and the Laplace transform of time $t$. The tildes indicate that the functions are represented in the Fourier-Laplace domain. $p_+^{eq}$, $p_0^{eq}$, and $p_-^{eq}$ denote the equilibrium probabilities of the states $\Gamma_+$, $\Gamma_0$, and $\Gamma_-$, respectively. A summation of the PDFs given in Eq. (S1) provides the PDF of multi-mode active matter, which is given by

$$\tilde{P}_M(w,s) \equiv \tilde{P}_-(w,s) + \tilde{P}_0(w,s) + \tilde{P}_+(w,s)$$
$$= \frac{\chi(w,s) + p_0^{eq} v_a^2 w^2}{(s+D_0w^2)(\chi(w,s)+v_a^2w^2)+2k_a v_a^2 w^2}. \quad (S2)$$

The denominator in Eq. (S2) is a cubic function of $z \equiv s + D_0 w^2$ as $z^3 + 2(k_0+k_a)z^2 + (k_0(k_0+2k_a)+v_a^2w^2)z + 2k_a v_a^2 w^2$. If we assume the roots of the cubic function as $-C_i(w)$ with $(i \in 1,2,3)$, the PDF can be rewritten as

$$\tilde{P}_M(w,s) = \frac{1}{s+D_0w^2} - 2k_a v_a^2 w^2 \left( \frac{1}{s+D_0w^2} + \frac{1}{k_0+2k_a} \right) \prod_{i=1}^{3} \frac{1}{s+D_0w^2+C_i(w)}$$
$$= \frac{1}{s+D_0w^2} - 2k_a v_a^2 w^2 \left( \frac{1}{s+D_0w^2} + \frac{1}{k_0+2k_a} \right) \sum_{i=1}^{3} \frac{1}{s+D_0w^2+C_i(w)} \prod_{j \neq i} \frac{1}{C_i(w)-C_j(w)}. \quad (S3)$$

Inverting $s$ in $\tilde{P}(w,s)$ generates the Fourier-domain PDF, written as

$$\hat{P}_M(w,t) = v_a^2 w^2 e^{-D_0 w^2 t} \left( \sum_{i=1}^{3} \left( \frac{2k_a}{C_i(w)} - \frac{2k_a}{k_0+2k_a} \right) e^{-C_i(w)t} \prod_{j \neq i} \frac{1}{C_i(w)-C_j(w)} \right). \quad (S4)$$

Eqs. (S2) and (S4) can both be used to derive the analytic solution for the mean square displacement (MSD) of the multi-mode model. One way is to use the second partial derivative of $\hat{P}_M(w,t)$, while the other way, which is an easier way to obtain the time-domain MSD, is to apply the inverse Laplace transform to the second partial derivative of $\tilde{P}_M(w,s)$, written as

$$\langle x^2(t) \rangle = \lim_{w \to 0} \left[ -\frac{\partial^2 \hat{P}_M(w,t)}{\partial w^2} \right] = \underset{s \to t}{L^{-1}} \left( \lim_{w \to 0} \left[ -\frac{\partial^2 \tilde{P}_M(w,s)}{\partial w^2} \right] \right), \tag{S5}$$

where $\underset{s \to t}{L^{-1}}$ denote the inverse Laplace transform of time.

The time-dependent MSD of this model is given in Eq. (6). The analytic solution for the fourth moment of displacement is obtained using the following equation:

$$\langle x^4(s) \rangle = \lim_{w \to 0} \left[ \frac{\partial^4 \tilde{P}_M(w,s)}{\partial w^4} \right] = \frac{24}{s^3} \left[ \begin{array}{l} D_0^2 + \dfrac{2v_a^2 k_a}{(s+k_0)(k_0+2k_a)} \times \\ \left( 3D_0 - \dfrac{1}{s+k_0}\left( k_0 D_0 - v_a^2 + \dfrac{v_a^2 k_0}{s+k_0+2k_a} \right) \right) \end{array} \right]. \tag{S6}$$

Application of the inverse Laplace transform to Eq. (S6) provides the fourth moment of displacement in the time domain, written as

$$\langle x^4(t) \rangle = 12(D_0 + D_a)^2 t^2 \\ + \frac{24 D_a^2 \tau_c^2}{(R+1)^2} \left( \begin{array}{l} R^5 e^{-(1+1/R)t/\tau_c} + \left(-R^2 + 3R - 3\right)(R+1)^3 e^{-t/\tau_c} \\ -\left(3R^3 - R^2 - 6R - 3\right) \\ +(R+1)^2 \left(R^2 - 1 + D_0/D_a\right) e^{-t/\tau_c} t/\tau_c \\ +(R+1)\left(R^2 - 2R - 2 - (R+1)D_0/D_a\right) t/\tau_c \end{array} \right) \tag{S7}$$

The second and fourth moments expressed in the time domain are combined to produce the non-Gaussian parameter, $\alpha(t)$, such as

$$\alpha_R(t) \equiv \frac{\langle x^4(t)\rangle}{3\langle x^2(t)\rangle^2} - 1 = \frac{1}{3}\kappa(t) - 1$$

$$= \frac{1}{(R+1)^4}\left(\frac{2D_a^{(0)}\tau_c}{\langle x^2(t)\rangle}\right)^2 \begin{pmatrix} -e^{-2t/\tau_c} - 4e^{-t/\tau_c} + 5 - 4e^{-t/\tau_c}\,t/\tau_c - 2t/\tau_c \\ +\left(-2e^{-2t/\tau_c} - 8e^{-t/\tau_c} + 10 - 8e^{-t/\tau_c}\,t/\tau_c - 4t/\tau_c\right)R \\ +\left(-e^{-2t/\tau_c} + 1 - 2e^{-t/\tau_c}\,t/\tau_c\right)R^2 \\ +\left(6e^{-t/\tau_c} - 6 + 4e^{-t/\tau_c}\,t/\tau_c + 2t/\tau_c\right)R^3 \\ +2\left(e^{-\frac{1}{R}t/\tau_c} - 1 + \frac{1}{R}t/\tau_c\right)e^{-t/\tau_c}R^5 \end{pmatrix}$$

with $\dfrac{\langle x^2(t)\rangle}{2D_a^{(0)}\tau_c} = \dfrac{D_0}{D_a^{(0)}}\dfrac{t}{\tau_c} + \dfrac{1}{(R+1)}\left(E^{-\frac{t}{\tau_c}} + \dfrac{t}{\tau_c} - 1\right)$,  (S8)

where $\kappa(t)$ denotes kurtosis and $D_a^{(0)} \equiv (R+1)D_a = \tau_c v_a^2$. On the log-scale time axis as shown in Fig. 3 (b), the relaxation time, $\tau_c$, shifts the $\alpha_R(t)$ curve as well as the MSD curve. In Fig. S1, $\alpha_R(t)$ in all ranges of $R$ and $t/\tau_c$ is analytically evaluated and plotted under the condition of $D_0/D_a^{(0)} = 0.1$, where the red lines are the two lines shown in Figs. 3(c) and 3(e) and the black line marks a border line switching from sub-Gaussian to super-Gaussian at a given $R$. Although $\alpha(t)$ depends on $D_0/D_a^{(0)}$, the border line is invariant on the change of the $D_0/D_a^{(0)}$ ratio in Eq. (S8). Thus, $\alpha_R(t)$ with $R$ less than $(1+\sqrt{5})/2$ is sub-Gaussian at all times, and $\alpha_R(t)$ with $R$ larger than 2 is always super-Gaussian at all times. When $(1+\sqrt{5})/2 < R < 2$, $P_M(x,t)$ can switch from sub-Gaussian to super-Gaussian over time, as shown in Figs. 3(e) and S1.

### B. Derivation of the second and fourth moments of displacement for the single-mode model

The single-mode model has two internal states, $\Gamma_+$ and $\Gamma_-$. Each internal state regulates the direction and speed of active matter, as explained in Fig. 1(a). Based on the initial conditions that the internal states are initially in equilibrium and the initial position of active matter is zero, two simultaneous equations are obtained from Eq. (3) by applying the Fourier transform and the Laplace transform to $P_i(x,t)$ with $i \in +, -$. The analytic solution of the simultaneous equations provides two PDFs in the Fourier-Laplace domain, written as

$$\begin{pmatrix} \tilde{P}_+(w,s) \\ \tilde{P}_-(w,s) \end{pmatrix} = \frac{1}{2} \frac{1}{v_a^2 w^2 + (s + D_0 w^2)(s + D_0 w^2 + 2k_a)} \begin{pmatrix} s + D_0 w^2 + 2k_a - iv_a w \\ s + D_0 w^2 + 2k_a + iv_a w \end{pmatrix}. \tag{S9}$$

The PDF of active matter for the single-mode model in Fourier-Laplace domain is written as

$$\begin{aligned} \tilde{P}_S(w,s) &\equiv \tilde{P}_-(w,s) + \tilde{P}_+(w,s) \\ &= \frac{s + D_0 w^2 + 2k_a}{v_a^2 w^2 + (s + D_0 w^2)(s + D_0 w^2 + 2k_a)}. \end{aligned} \tag{S10}$$

Application of the inverse Laplace transform to Eq. (S10) generates the PDF of active matter represented in the Fourier domain as

$$\hat{P}_S(w,t) = e^{-t(k_a + D_0 w^2)} \left( \cosh(t\Lambda) + \frac{k_a}{\Lambda} \sinh(t\Lambda) \right) \quad \text{with} \quad \Lambda \equiv \sqrt{k_a^2 - v_a^2 w^2}. \tag{S11}$$

From this function, the time-dependent second and fourth moments of displacement are simple to obtain. The time-dependent MSD of this model is given in Eq. (6). The fourth moment of displacement is also evaluated from the PDF as

$$\langle x^4(t) \rangle = 12 \tau_c^2 \left( (D_0 + D_a)^2 \frac{t^2}{\tau_c^2} + 2D_a \left( (D_a - D_0)(1 - e^{-t/\tau_c}) \frac{t}{\tau_c} + 3D_a \left( 1 - e^{-t/\tau_c} - \frac{t}{\tau_c} \right) \right) \right), \tag{S12}$$

where $D_a = \tau_c v_a^2$. The second and fourth moments expressed in the time domain are combined to produce the non-Gaussian parameter, $\alpha(t)$, such as

$$\alpha_0(t) = \left( \frac{2D_a \tau_c}{\langle x^2(t) \rangle} \right)^2 \left( 5 - e^{-2t/\tau_c} - 4e^{-t/\tau_c} - 4e^{-t/\tau_c} t/\tau_c - 2t/\tau_c \right) = \left( \frac{2D_a \tau_c}{\langle x^2(t) \rangle} \right)^2 \beta(t)$$

with $\dfrac{\langle x^2(t) \rangle}{2 D_a \tau_c} = \dfrac{D_0}{D_a} \dfrac{t}{\tau_c} + \left( E^{-\frac{t}{\tau_c}} + \dfrac{t}{\tau_c} - 1 \right)$ and

$$\beta(t) \equiv 5 - e^{-2t/\tau_c} - 4e^{-t/\tau_c} - 4e^{-t/\tau_c} t/\tau_c - 2t/\tau_c. \tag{S13}$$

In Eq. (S13), $\alpha_0(t)$ is less than or equal to zero because $\beta(t) \leq 0$ in all time ranges, and the equation is equal to $\alpha_R(t)$ of the multi-mode model at the small $R$ limit as $\alpha_0(t) = \lim_{R \to 0} \alpha_R(t)$.

### C. Probability density function of displacement at two limiting time scales

The diffusion dynamics of the models is highly dependent on the relaxation time, $\tau_c$, of the velocity, $v_s(\Gamma)$. At short times ($t \ll \tau_c$), a given active matter maintains its direction and magnitude of velocity, and each unrelaxed velocity produces three individual peaks in the PDF of displacement. The PDF of displacement at short times is derived from Eq. (S1), which is written as

$$\begin{pmatrix} \tilde{P}_{short,+}(w,s) \\ \tilde{P}_{short,0}(w,s) \\ \tilde{P}_{short,-}(w,s) \end{pmatrix} = \begin{pmatrix} p_+^{eq}(s+D_0 w^2 + i v_a w)^{-1} \\ p_0^{eq}(s+D_0 w^2)^{-1} \\ p_-^{eq}(s+D_0 w^2 - i v_a w)^{-1} \end{pmatrix}. \tag{S14}$$

$P_M(x,t)$ at short times is written as

$$P_{M,short}(x,t) = \frac{1}{\sqrt{4\pi D_0 t}} \left( p_-^{eq} e^{-\frac{(x+v_a t)^2}{4D_0 t}} + p_0^{eq} e^{-\frac{x^2}{4D_0 t}} + p_+^{eq} e^{-\frac{(x-v_a t)^2}{4D_0 t}} \right), \tag{S15}$$

where the distribution is Gaussian with a variance of $2D_0 t$. The three peaks in $P_{M,short}(x,t)$ are approximated as a single Gaussian function with a small variance at very short times ($t \ll 2D_0/v_a^2$) and gradually separate as time increases.

At long times ($t \gg \tau_c$), the peaks for individual $v_s(\Gamma)$s are again intermingled into a single Gaussian and follow the distribution, written as

$$\begin{pmatrix} \tilde{P}_{long,+}(w,s) \\ \tilde{P}_{long,0}(w,s) \\ \tilde{P}_{long,-}(w,s) \end{pmatrix} = \frac{1}{s+D_{eff}w^2} \begin{pmatrix} p_+^{eq}(1-iv_a w/k_0) \\ p_0^{eq}(1+v_a^2 w^2/k_0(k_0+2k_a)) \\ p_-^{eq}(1+iv_a w/k_0) \end{pmatrix}, \tag{S16}$$

where $D_{eff}$ is equal to $D_0 + D_a$. The PDF $P_M(x,t)$ at long times is written as

$$P_{M,long}(x,t) = \frac{1}{\sqrt{4\pi D_{eff}t}} e^{-\frac{x^2}{4D_{eff}t}} \left( 1 + \frac{p_0^{eq} D_a \tau_c R}{2D_{eff}t} \left( 1 - \frac{x^2}{2D_{eff}t} \right) \right). \tag{S17}$$

In Eq. (S17), deviation from the Gaussian distribution is proportional to $p_0^{eq} D_a \tau_c R / 2D_{eff} t$, where $p_0^{eq} D_a / D_{eff}$ is always less than 1, and $t$ is larger than $\tau_c$ as $t/\tau_c \gg 1$ at long times. Therefore, $p_0^{eq} D_a \tau_c R / 2D_{eff} t$ is much smaller than 1 if $R$ is finite at sufficiently long times. Thus, the PDF approaches the Gaussian distribution at sufficiently long times, which is in accordance with the Gaussian central limit theorem.

In summary, the PDF of displacement approaches the delta function at very short times, because self-propelled velocity is much weaker than the velocity caused by the random fluctuating force. As a result of the contribution of random fluctuation being dissipated, the delta function is split into individual peaks related to the velocity of each internal state, whereas the self-propelled velocity shows no variation. Eq. (S15) explains the two different functional forms of the PDF. At long times, the PDF approaches the dispersed Gaussian distribution because the variance for each distribution is proportional to time.

### D. Mean velocity distribution and stationary distribution

The mean velocity, $\bar{v}(t)$, is defined by $\bar{v}(t) \equiv x(t)/t$. The mean velocity distribution, $f_M(\bar{v},t)$, is directly obtained from the PDF of displacement with a proper normalization constant. At short times ($t \ll \tau_c$), the mean velocity distribution, $f_M(\bar{v},t)$, is written as

$$f_{M,short}(\bar{v},t) = \frac{1}{\sqrt{4\pi D_0/t}} \left( p_-^{eq} e^{-\frac{(\bar{v}+v_a)^2}{4D_0/t}} + p_0^{eq} e^{-\frac{\bar{v}^2}{4D_0/t}} + p_+^{eq} e^{-\frac{(\bar{v}-v_a)^2}{4D_0/t}} \right). \tag{S18}$$

where the distribution is Gaussian with a variance of $2D_0/t$. Because the variance is inversely proportional to $t$, the broadness of the individual peaks in Eq. (S18) shows the opposite pattern compared to the individual peaks in $P_M(x,t)$, which appears in Eq. (S15). At long times ($t \gg \tau_c$), $f_M(\bar{v},t)$ is written as

$$f_{M,long}(\bar{v},t) = \frac{1}{\sqrt{4\pi D_{eff}/t}} e^{-\frac{\bar{v}^2}{4D_{eff}/t}} \left( 1 + \frac{p_0^{eq} D_a \tau_c R}{2D_{eff} t} \left( 1 - \frac{\bar{v}^2}{2D_{eff}/t} \right) \right).$$

(S19)

$f_{M,long}(\bar{v},t)$ approaches the delta function as time increases.

The variance of the individual peaks in $P_M(x,t)$ is proportional to $t$ at both short and long times. If we define a new variable, $q(\equiv x/\sqrt{t})$, then the stationary distribution can be obtained at two the time-limiting cases. At short times ($t \ll \tau_c$), the stationary distribution, $g_M(q,t)$, is written as

$$g_{M,short}(q,t) = \frac{1}{\sqrt{4\pi D_0}} \left( p_- e^{-\frac{(q+v_a\sqrt{t})^2}{4D_0}} + p_0 e^{-\frac{q^2}{4D_0}} + p_+ e^{-\frac{(q-v_a\sqrt{t})^2}{4D_0}} \right). \tag{S20}$$

The variance of the distribution, $g_M(q,t)$, at short times is equal to $2D_0$ and does not vary with time, however, the interval between the peaks does gradually increase as time increases. At long times ($t \gg \tau_c$), $g_M(q,t)$ is written as

$$g_{M,long}(q,t) = \frac{1}{\sqrt{4\pi D_{eff}}} e^{-\frac{q^2}{4D_{eff}}} \left(1 + \frac{p_0^{eq} D_a \tau_c R}{2 D_{eff} t}\left(1 - \frac{q^2}{2 D_{eff}}\right)\right). \tag{S21}$$

$g_{M,long}(q,t)$ converges to the Gaussian distribution with a variance of $2 D_{eff}$.

### E. Dynamics of the multi-mode model

For the multi-mode model, a given active matter is operated by a variable composed of three discrete states: $\Gamma_+$, $\Gamma_0$, and $\Gamma_-$. If the active matter with a $\Gamma_i$ state is located in the infinitesimal area $dx$, then the probability of finding the active matter can be written as $\rho(\Gamma_i, x, t) dx$. The PDF satisfies the conservation law, written as

$$\int_{-\infty}^{\infty} dx \sum_i^3 \rho(\Gamma_i, x, t) = 1. \tag{S22}$$

From the conservation law, the continuity equation for the PDF is written as

$$\frac{\partial}{\partial t} \rho(\Gamma_i, x, t) = -\frac{\partial}{\partial x}(\dot{x} \rho(\Gamma_i, x, t)) + \sum_{j \neq i}^3 \left[-K_{i \to j} \rho(\Gamma_i, x, t) + K_{j \to i} \rho(\Gamma_j, x, t)\right],$$

where $\dot{x}$ and $K_{i \to j}$ denote the time derivative of $x$ and the rate constant from a state $\Gamma_i$ to $\Gamma_j$. Here, we consider the motion of active matter in an overdamped environment where acceleration is zero. The time derivative of an active matter position is written as

$$\frac{dx}{dt} = v(\Gamma) + \gamma^{-1} \cdot \xi(t), \tag{S23}$$

where $\xi(t)$ represents the random fluctuating force modeled as Gaussian white noise. The ensemble average of $\rho(\Gamma_i, x, t)$ over the Gaussian white noise gives the observed PDF $P(\Gamma_i, x, t)$ [1]. The application of the cumulant expansion gives Eqs. (3) and (4), where we set $P(\Gamma_i, x, t)$ to $P_i(x, t)$ for the sake of simplicity. Internal-state-dependent velocity, $v(\Gamma)$, depends on the state as $v(\Gamma_+) = v_a$, $v(\Gamma_0) = 0$, and $v(\Gamma_-) = -v_a$.

### F. Convergence of $P_M(x,t)$ to $P_S(x,t)$ at the small $R$ limit

The population ratio, $R(\equiv p_0^{eq}/(p_+^{eq} + p_-^{eq}) = \tau_0/\tau_a = k_0/2k_a)$, of the passive state to the active state modulates the shape of the probability density of the active matter, $P_M(x,t)$, in the multi-mode model. Applying $k_0 = \tau_c^{-1}$ and $2k_a = R^{-1}\tau_c^{-1}$ to Eq. (S2) produces $\tilde{P}_M(w,s)$, written as

$$\tilde{P}_M(w,s) = \frac{\left(s+D_0w^2+\tau_c^{-1}\right)\left(s+D_0w^2+\tau_c^{-1}+R^{-1}\tau_c^{-1}\right)+v_a^2w^2R/(R+1)}{\left(s+D_0w^2\right)\left(\left(s+D_0w^2+\tau_c^{-1}\right)\left(s+D_0w^2+\tau_c^{-1}+R^{-1}\tau_c^{-1}\right)+v_a^2w^2\right)+R^{-1}\tau_c^{-1}v_a^2w^2}$$

$$= \frac{\left(s+D_0w^2+\tau_c^{-1}\right)\left(Rs+RD_0w^2+R\tau_c^{-1}+\tau_c^{-1}\right)+v_a^2w^2R^2/(R+1)}{\left(s+D_0w^2\right)\left(\left(s+D_0w^2+\tau_c^{-1}\right)\left(Rs+RD_0w^2+R\tau_c^{-1}+\tau_c^{-1}\right)+Rv_a^2w^2\right)+\tau_c^{-1}v_a^2w^2}.$$

(S24)

In the limit of $R \to 0$, $\tilde{P}_M(w,s)$ is written as

$$\lim_{R \to 0} \tilde{P}_M(w,s) = \frac{s+D_0w^2+\tau_c^{-1}}{\left(s+D_0w^2\right)\left(s+D_0w^2+\tau_c^{-1}\right)+v_a^2w^2}.$$

(S25)

The relaxation time of the single-mode model is $\tau_c = (2k_a)^{-1}$. Applying $2k_a = \tau_c^{-1}$ to Eq. (S10) produces $\tilde{P}_S(w,s)$ as

$$\tilde{P}_S(w,s) = \frac{s+D_0w^2+\tau_c^{-1}}{\left(s+D_0w^2\right)\left(s+D_0w^2+\tau_c^{-1}\right)+v_a^2w^2}.$$

(S26)

$\tilde{P}_S(w,s)$ is the same as $\tilde{P}_M(w,s)$ in the small $R$ limit.

### G. Relaxation time of the two solvable models

In the high friction regime, where we can safely neglect the inertial term in the Langevin equation, the velocity, $\dot{x}(t)$, of active matter with a friction constant, $\gamma$, can be written as the sum of two components:

$$\dot{x}(t) = v_s(\Gamma(t)) + \gamma^{-1}\xi(t),$$

(S27)

where $v_s(\Gamma)$ and $\gamma^{-1}\xi(t)$ represent the velocity component of a self-propelled, ballistic motion, which is dependent on the internal state, $\Gamma$, and the velocity component caused by the random fluctuating force. If we assume that the initial position of active matter is zero, then the time integration of Eq. (S27) produces the time-dependent position, written as

$$x(t) = \int_0^t \left(v_s(\Gamma(\tau)) + \gamma^{-1}\xi(\tau)\right)d\tau.$$

(S28)

From Eq. (S28), the MSD of the active matter can be evaluated from the velocity correlation function, written as

$$\langle x^2(t) \rangle = \int_0^t d\tau_2 \int_0^t d\tau_1 \left[\langle v_s(\Gamma(\tau_2))\cdot v_s(\Gamma(\tau_1))\rangle + \frac{1}{\gamma^2}\langle \xi(\tau_2)\cdot\xi(\tau_1)\rangle\right],$$

$$= 2D_0 t + 2\int_0^t d\tau (t-\tau)\langle v_s(\tau)\cdot v_s(0)\rangle$$

(S29)

where we denote $v_s(\Gamma(\tau))$ in short as $v_s(\tau)$. By comparing Eq. (S29) with Eq. (6), we obtain the following equation:

$$\int_0^t d\tau (t-\tau)\langle v_s(\tau) \cdot v_s(0)\rangle = D_a \tau_c \left(e^{-t/\tau_c} - 1 + t/\tau_c\right). \tag{S30}$$

Because $D_a$ is equal to $\tau_c \langle v_s^2 \rangle$, the second derivative of each side of Eq. (S30) provides the normalized time correlation function of velocity, $\phi_{v_s}(t)$, as

$$\phi_{v_s}(t) = \langle v_s(t) v_s(0)\rangle / \langle v_s^2 \rangle = e^{-t/\tau_c}, \tag{S31}$$

where $\tau_c$ is given by $(2k_a)^{-1}$ for the single-mode model and $k_0^{-1}$ for the multi-mode model.

## H. General model

In general, a given active matter moves in a multidimensional space, $d$, and its random fluctuating force has a finite relaxation time, $\tau_p$. To obtain analytic solutions for this general model, the velocity of active matter corresponding to Eq. (1) is generalized to

$$\dot{\mathbf{r}}(t) = \mathbf{v}_s(\Gamma(t)) + \gamma^{-1}\boldsymbol{\xi}(t), \tag{S32}$$

where each bold symbol denotes the $d$-dimensional vector corresponding to each scalar quantity in Eq. (1). The integration of each side of Eq. (S32) from 0 to $t$ produces the time-dependent position, $\mathbf{r}(t)$, written as

$$\mathbf{r}(t) = \int_0^t \left(\mathbf{v}_s(\Gamma(\tau)) + \gamma^{-1}\boldsymbol{\xi}(\tau)\right) d\tau, \tag{S33}$$

where we assume the initial position is zero. From Eq. (S33), the MSD is written as

$$\begin{aligned}\langle |\mathbf{r}(t)|^2 \rangle &= 2\int_0^t d\tau(t-\tau)\left[\gamma^{-2}\langle \boldsymbol{\xi}(\tau)\cdot\boldsymbol{\xi}(0)\rangle + \langle \mathbf{v}_s(\tau)\cdot\mathbf{v}_s(0)\rangle\right] \\ &= 2d\int_0^t d\tau(t-\tau)\left[D_0 \tau_p^{-1}\phi_\xi(\tau) + D_a \tau_c^{-1}\phi_{v_s}(\tau)\right] \\ &= \langle |\mathbf{r}_\xi(t)|^2\rangle + \langle |r_{v_s}(t)|^2\rangle\end{aligned} \tag{S34}$$

where $\phi_\xi(t)$ denotes the normalized time correlation function, $\langle \boldsymbol{\xi}(t)\cdot\boldsymbol{\xi}(0)\rangle/\langle \boldsymbol{\xi}(0)^2\rangle$, of the random fluctuating force, $\boldsymbol{\xi}(t)$, and the relaxation time, $\tau_p$, is defined as $\tau_p \equiv \int_0^\infty dt\,\phi_\xi(t)$. Here, the diffusion coefficient for passive motion is defined by $D_0 = d^{-1}\gamma^{-2}\int_0^\infty dt\,\langle \boldsymbol{\xi}(t)\cdot\boldsymbol{\xi}(0)\rangle$, and the diffusion coefficient for self-propelled motion is defined by $D_a = d^{-1}\int_0^\infty dt\,\langle \mathbf{v}_s(t)\cdot\mathbf{v}_s(0)\rangle$. The MSD consists of two independent movements from the diffusive mode and the self-active mode. The diffusive mode contribution to the MSD is defined as $\left\langle \left|\mathbf{r}_\xi(t)\right|^2\right\rangle \equiv 2dD_0\tau_p^{-1}\int_0^t d\tau(t-\tau)\phi_\xi(\tau)$, while the self-propelled mode contribution is defined as $\left\langle \left|\mathbf{r}_{\mathbf{v}_s}(t)\right|^2\right\rangle \equiv 2dD_a\tau_c^{-1}\int_0^t d\tau(t-\tau)\phi_{\mathbf{v}_s}(\tau)$. If we assume that the distribution of $\xi(t)$ is Gaussian, then the analytic solutions for the fourth moment of displacement can be written as

$$\left\langle \left|\mathbf{r}(t)\right|^4\right\rangle = \left(1+2d^{-1}\right)\left\langle \left|\mathbf{r}_\xi(t)\right|^2\right\rangle^2 + 2\left(1+2d^{-1}\right)\left\langle \left|\mathbf{r}_\xi(t)\right|^2\right\rangle\left\langle \left|\mathbf{r}_{\mathbf{v}_s}(t)\right|^2\right\rangle + \left\langle \left|\mathbf{r}_{\mathbf{v}_s}(t)\right|^4\right\rangle, \quad (S35)$$

where $\left\langle \left|\mathbf{r}_{\mathbf{v}_s}(t)\right|^4\right\rangle \equiv 4!\int_0^t dt_4 \int_0^{t_4} dt_3 \int_0^{t_3} dt_2 \int_0^{t_2} dt_1 \left\langle \mathbf{v}_s(t_4)\cdot\mathbf{v}_s(t_3)\mathbf{v}_s(t_2)\cdot\mathbf{v}_s(t_1)\right\rangle$. The non-Gaussian parameter for the general model is written as

$$\alpha_R(t) \equiv \frac{d}{d+2}\frac{\left\langle \left|\mathbf{r}(t)\right|^4\right\rangle}{\left\langle \left|\mathbf{r}(t)\right|^2\right\rangle^2} - 1 = \frac{\left\langle \left|\mathbf{r}_{\mathbf{v}_s}(t)\right|^2\right\rangle^2}{\left\langle \left|\mathbf{r}(t)\right|^2\right\rangle^2}\alpha_{\mathbf{v}_s}(t)$$

$$\text{with } \alpha_{\mathbf{v}_s}(t) \equiv \frac{d}{d+2}\frac{\left\langle \left|\mathbf{r}_{\mathbf{v}_s}(t)\right|^4\right\rangle}{\left\langle \left|\mathbf{r}_{\mathbf{v}_s}(t)\right|^2\right\rangle^2} - 1. \quad (S36)$$

At short times, $\alpha_R(t)$ approaches zero because $\left\langle \left|\mathbf{r}(t)\right|^2\right\rangle^2 \gg \left\langle \left|\mathbf{r}_{\mathbf{v}_s}(t)\right|^2\right\rangle^2$. At long times, $\alpha_R(t)$ also approaches zero because $\alpha_{\mathbf{v}_s}(t)$ approaches zero.

### I. Time correlation functions for two solvable models

Time correlation functions of the velocity component, $v_s$, of self-propelled motion are used to calculate $D_a$ and $\alpha(t)$ as well as the second and fourth moments of displacement. In our model, because $v_s$ is only dependent on the internal state, we analytically obtain the time evolution of internal state probabilities as

$$\begin{pmatrix} P_+(t) \\ P_-(t) \end{pmatrix} = \mathbf{G}\begin{pmatrix} P_+(0) \\ P_-(0) \end{pmatrix} \text{ with } \mathbf{G} = E^{-k_a t}\begin{pmatrix} \cosh(k_a t) & \sinh(k_a t) \\ \sinh(k_a t) & \cosh(k_a t) \end{pmatrix} \quad (S37)$$

for the single-mode model and

$$\begin{pmatrix} P_+(t) \\ P_0(t) \\ P_-(t) \end{pmatrix} = \mathbf{G}\begin{pmatrix} P_+(0) \\ P_0(0) \\ P_-(0) \end{pmatrix} \text{ with}$$

$$\mathbf{G} = \frac{1}{k_0 + 2k_a} \begin{pmatrix} k_a + \frac{k_0 e^{-(k_0+2k_a)t} + (k_0+2k_a)e^{-k_0 t}}{2} & k_a - k_a e^{-(k_0+2k_a)t} & k_a + \frac{k_0 e^{-(k_0+2k_a)t} - (k_0+2k_a)e^{-k_0 t}}{2} \\ k_0 - k_0 e^{-k_0-2k_a} & k_0 + 2k_a e^{-(k_0+2k_a)t} & k_0 - k_0 e^{-k_0-2k_a} \\ k_a + \frac{k_0 e^{-(k_0+2k_a)t} - (k_0+2k_a)e^{-k_0 t}}{2} & k_a - k_a e^{-(k_0+2k_a)t} & k_a + \frac{k_0 e^{-(k_0+2k_a)t} + (k_0+2k_a)e^{-k_0 t}}{2} \end{pmatrix}$$

(S38)

for the multi-mode model. We obtain the velocity autocorrelation function, $\langle \mathbf{v}_s(t) \cdot \mathbf{v}_s(0) \rangle$, through the following equation:

$$\langle v_s(t) v_s(0) \rangle = \sum_{i \in +,-} \sum_{j \in +,-} v_s(\Gamma_j) v_s(\Gamma_i) \mathbf{G}(\Gamma_j, t | \Gamma_i, 0) P(\Gamma_i, 0),$$
(S39)

where $\mathbf{G}(\Gamma_j, t | \Gamma_i, 0)$ denotes a transition matrix from $\Gamma_i$ to $\Gamma_j$ after $t$ time passing. The transition matrices are shown in Eq. (S37) for the single-mode model and in Eq. (S38) for the multi-mode model. The calculation results of $\langle v_s(t) v_s(0) \rangle$ are $v_a^2 e^{-2k_a t}$ for the single-mode model and $p_a v_a^2 e^{-k_0 t}$ for the multi-mode model, which coincides with Eq. (S31). By calculating the MSD through the application of the calculated velocity autocorrelation function to Eq. (S29), we obtain Eq. (6). The four-time velocity autocorrelation function, $\langle v_s(t_4) v_s(t_3) v_s(t_2) v_s(t_1) \rangle$, is obtained by the following equation, written as

$$\begin{aligned} &\langle v_s(t_4) v_s(t_3) v_s(t_2) v_s(t_1) \rangle \\ &= \sum_{i \in +,-} \sum_{j \in +,-} \sum_{k \in +,-} \sum_{l \in +,-} v_s(\Gamma_l) v_s(\Gamma_k) v_s(\Gamma_j) v_s(\Gamma_i) \\ &\quad \times \mathbf{G}(\Gamma_l, t_4 | \Gamma_k, t_3) \mathbf{G}(\Gamma_k, t_3 | \Gamma_j, t_2) \mathbf{G}(\Gamma_j, t_2 | \Gamma_i, t_1) P(\Gamma_i, t_1) \end{aligned}$$

The calculation results of $\langle v_s(t_4) v_s(t_3) v_s(t_2) v_s(t_1) \rangle$ are $v_a^4 e^{-2k_a(t_4-t_3+t_2-t_1)}$ for the single-mode model and $p_a^2 v_a^4 \left( e^{k_0(t_3-t_2)} + R e^{-2k_a(t_3-t_2)} \right) e^{-k_0(t_4-t_1)}$ for the multi-mode model. The four-time velocity autocorrelation functions can be used to generate $\langle |\mathbf{r}_{\mathbf{v}_s}(t)|^4 \rangle$ in Eq. (S35), and their results are equal to the fourth moment of displacement which are written in Eqs. (S7) and (S12).

### J. Derivation of short time mean velocity distribution from Eq. (1)

In our model, the velocity of active matter consists of two components in Eq. (1). If we assume that the two components are independent, then the mean velocity distribution is written as

$$f(\bar{v}, t) = \int dv_s dv_\xi \delta(\bar{v} - (v_s + v_\xi)) f_s(v_s, t) f_\xi(v_\xi, t),$$
(S40)

where $v_s$ and $f_s(v_s,t)$ respectively denote the velocity component caused by self-propelled motion and its distribution function; $v_\xi$ and $f_\xi(v_\xi,t)$ denote the velocity component due to the random fluctuating force and its distribution function. At short times, the velocity component, $\gamma^{-1}\xi(t)$, caused by the random fluctuating force is already relaxed and follows a Gaussian distribution with a variance of $2D_0/t$, whereas the self-propelled motion approximately maintains its direction. The two distribution functions at short times can be written as

$$f_\xi(v_\xi,t) = e^{-\frac{v_\xi^2}{4D_0/t}} \Big/ \sqrt{4\pi D_0/t} \quad \text{and} \quad f_s(v_s,t) = \sum_{i\in\Gamma} p_i^{eq} \delta(v_s - v_i). \tag{S41}$$

By applying Eq. (S41) to Eq. (S40), the mean velocity distribution function at short times can be rewritten as

$$f_{short}(\bar{v},t) = \frac{1}{2\pi} \int_{-\infty}^{\infty} du \int dv_s dv_\xi \, e^{iu(\bar{v}-(v_s+v_\xi))} \sum_{i\in\Gamma} p_i^{eq} \delta(v_s - v_i) f_\xi(v_\xi,t)$$

$$= \frac{1}{2\pi} \sum_{i\in\Gamma} p_i^{eq} \int_{-\infty}^{\infty} du \, e^{iu(\bar{v}-v_i)} \int dv_\xi \, e^{-iuv_\xi} f_\xi(v_\xi,t)$$

$$= \sum_{i\in\Gamma} p_i^{eq} \int dv_\xi \, f_\xi(v_\xi,t) \delta(\bar{v}-v_i-v_\xi) = \sum_{i\in\Gamma} p_i^{eq} f_\xi(\bar{v}-v_i,t)$$

$$= \sum_{i\in\Gamma} p_i^{eq} \frac{1}{\sqrt{4\pi D_0/t}} e^{-\frac{(\bar{v}-v_i)^2}{4D_0/t}}. \tag{S42}$$

Eq. (S42) is equivalent to the mean velocity distribution for the multi-mode model at short times, which is shown in Eq. (S18).

### K. Stochastic simulation method

Our stochastic simulation method consists of both the Brownian dynamics for the time evolution of an active matter position and the Gillespie method for the stochastic transition between internal states [2,3]. For the Brownian dynamics, we numerically integrate Eq. (S23) as

$$x(t+\Delta t) = x(t) + v(\Gamma) \cdot \Delta t + \sqrt{2D_0 \cdot \Delta t} \cdot \xi'(t), \tag{S43}$$

where $x(t)$, $\Delta t$, and $\xi'(t)$ denote the active matter position at time $t$, the size of the time step, and the Gaussian random number with $G(0,1)$, respectively [3]. For the Gillespie method, we assume that the transitions between internal states for the multi-mode model are forbidden, except those transitions described by the following four unimolecular reactions:

$\Gamma_+ \xrightarrow{K_{+\to 0}=k_0} \Gamma_0$, $\Gamma_- \xrightarrow{K_{-\to 0}=k_0} \Gamma_0$, $\Gamma_0 \xrightarrow{K_{0\to +}=k_a} \Gamma_+$, and $\Gamma_0 \xrightarrow{K_{0\to -}=k_a} \Gamma_-$ [2]. The reaction constants for the forbidden transitions are set equal to zero. Our stochastic simulations proceed as follows:

1. Randomly choose an internal state of active matter based on the equilibrium population between states and set the initial position equal to zero. Set the selected state to the current state, $\Gamma_c$.

2. Based on the current state, calculate the waiting time for a reaction using the equation, $\tau = -\ln(RN) / \sum_{j\neq c} K_{c\to j}$, where $RN$ denotes an evenly distributed random number between 0 and 1, because concentration of the selected state is 1 and the concentration for the other states is zero. Only the $\Gamma_0$ state has two reaction paths with equal probability, and the other states have only one path for state transition.

3. Until the waiting time $\tau$ is over, evolve the time-dependent position using Eq. (S43) with the state-dependent velocity $v(\Gamma_c)$ and a given time interval $\Delta t$.

4. After finishing the time evolution in step 3, change the current state to the state determined by the transition in step 2. Return to step 2 when the elapsed time of the trajectory is less than the time limit of the trajectory.

5. Return to step 1 until sufficient trajectories are collected. We use 500,000 trajectories to obtain the velocity distributions and the second and fourth moments of the displacement distributions.

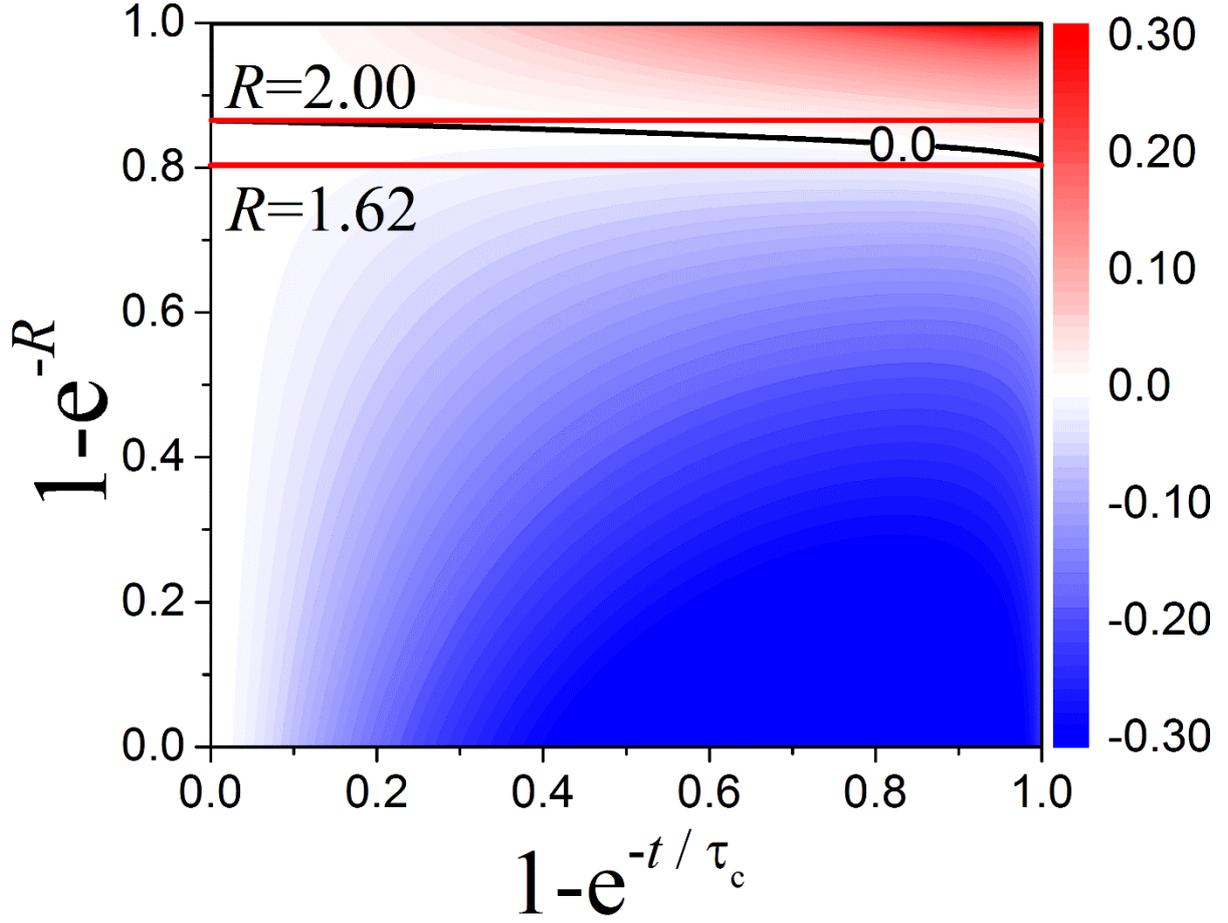

**Figure S1.** Non-Gaussian parameter. $\alpha_M(t)$ in all ranges of $R$ and $t/\tau_c$ are plotted when $D_0/D_a^{(0)}$ is equal to 0.1. The distribution of displacement is sub-Gaussian (super-Gaussian) in all time ranges if $R<1.62$ ($R>2.00$), as shown in Fig. 3(d). The two horizontal red lines represent $\alpha_M(t)$ for the two critical values of $R$: 1.62 and 2. In the range $1.62 < R < 2.00$, the displacement distribution switches from sub-Gaussian to super-Gaussian along the time axis. The boundary between the sub-Gaussian and the super-Gaussian distribution is represented by the black line.